\documentclass[12pt,preprint,tighten,iop]{aastex}
\usepackage[english]{babel}
\usepackage[]{times, graphicx}
\newcommand{\pref}{\protect\ref}

\newcommand{\hinode}{{\em Hinode{}}}

\newcommand{\degree}{${^\circ{}}$}

\begin{document}
\shorttitle{Overlap and Interaction}
\shortauthors{S.W.\ McIntosh and R.J.\ Leamon}

\title{On Magnetic Activity Band Overlap, Interaction, and the Formation of Complex Solar Active Regions}

\author{Scott W. McIntosh\altaffilmark{1,2}, Robert J. Leamon\altaffilmark{3}}

\altaffiltext{1}{High Altitude Observatory, National Center for Atmospheric Research, P.O. Box 3000, Boulder, CO 80307}
\altaffiltext{2}{School of Mathematics and Statistics, University of St Andrews, St Andrews, Fife, KY16 9SS, UK}
\altaffiltext{3}{Department of Physics, Montana State University, Bozeman, MT 59717}

\begin{abstract}
Recent work has revealed an phenomenological picture of the how the $\sim$11-year sunspot cycle of Sun arises. The production and destruction of sunspots is a consequence of the latitudinal-temporal overlap and interaction of the toroidal magnetic flux systems that belong to the 22-year magnetic activity cycle and are rooted deep in the Sun's convective interior. We present a conceptually simple extension of this work, presenting a hypothesis on how complex active regions can form as a direct consequence of the intra- and extra-hemispheric interaction taking place in the solar interior. Furthermore, during specific portions of the sunspot cycle we anticipate that those complex active regions may be particular susceptible to profoundly catastrophic breakdown---producing flares and coronal mass ejections of most severe magnitude.
\end{abstract}


\section{Introduction}\label{intro}
Recent work has exploited a novel analysis method for line-of-sight (LOS) magnetograms called the ``Magnetic Range of Interaction''  \citep{2007ApJ...654..650M} to illustrate the presence of the elusive giant convective scale \citep[e.g.,][]{1998Natur.394..653B,2013Sci...342.1217H} in the photosphere and to present markers of that scale \citep{McIntosh2014a}. The most familiar of these markers to the reader may be the (ubiquitous) coronal brightpoint \citep[e.g.,][]{1974ApJ...189L..93G}, although their potential association with the deep convective interior may have gone largely unnoticed. 

Following that discovery, 
\citeauthor{McIntosh2014b} (\citeyear{McIntosh2014b}; hereafter Paper~I) 
presented the temporal evolution of the markers of the giant scale and coronal brightpoints. They deduced the landmarks and phases of the sunspot cycle: the ascending phase, solar maximum, the declining phase and solar minimum arise as a result of the latitudinal-temporal interaction of (toroidal) magnetic flux systems (or what we will refer to hereafter as ``activity bands'') that belong to the 22-year magnetic activity cycle. In short, the appearance of the sunspot cycle could be explained in terms of the destructive ``interference'' of these oppositely signed activity bands as they migrate equatorward. 
 
\begin{figure}
\epsscale{1.00}
\plotone{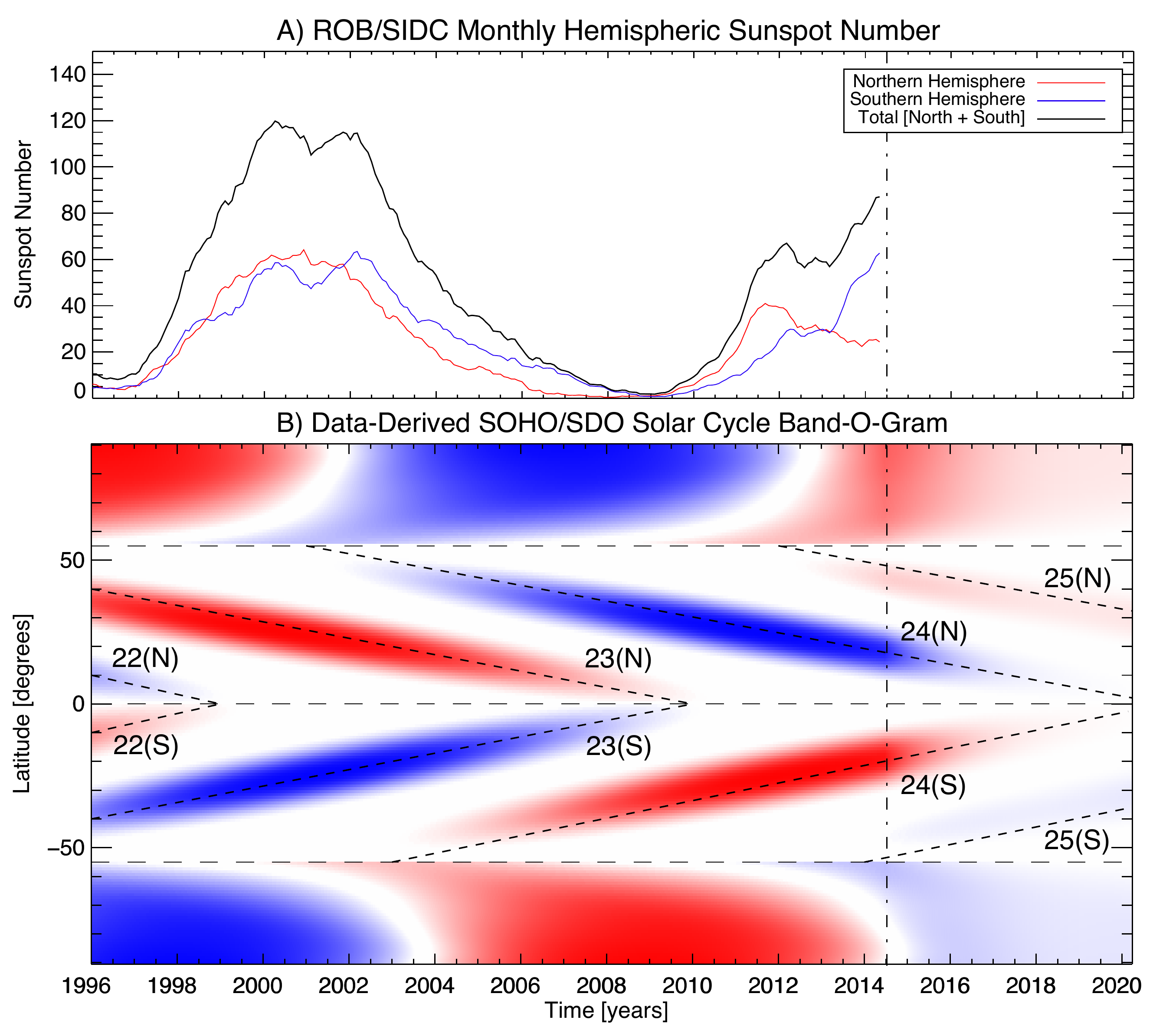}
\caption{Panel A shows the total and hemispheric sunspot numbers for cycles 23 and 24 as black, red and blue solid lines, respectively. Panel B shows the data-determined schematic of the migrating activity bands that belong to the 22-year magnetic activity cycle from \citet{McIntosh2014b}. The interplay of these bands gave rise to sunspot cycles 22, 23, and 24 as indicated. In each case the color of the band reflects its polarity (red--positive; blue--negative). The dashed horizontal lines drawn indicate the equator and lines of the 55th parallel in each hemisphere while the vertical dot-dashed lines indicate the current time. The reader will notice that, in addition to the cycle 25 bands appearance between 2012 and 2014, we have attempted to indicate the progression of the activity bands into the future as outlined in \citet{McIntosh2014b}. This projection doesn't carry information about the strength of activity on the bands in the future only their migratory paths. \label{f1}}
\end{figure}
 
The main result of Paper~I is illustrated in panel B of Fig.~\pref{f1}. In short, the temporal and latitudinal interaction of the oppositely-signed activity bands in each hemisphere, and across the equator, modulate the occurrence of sunspots (on the low-latitude pair). We noticed that the activity bands of same sign appear at high latitudes ($\sim$55\degree) every 22 years and migrate equatorward, taking approximately 19 years to reach the equator. The low-latitude pair of bands abruptly ``terminate'' at the equator. This equatorial termination, or cancellation, signals the end of one sunspot cycle and leaves only the higher-latitude band in each hemisphere. For example, the cycle 23 sunspots did not appear in grow in abundance (or area) until the cycle 22 bands had terminated in mid-1997, similarly in early 2011 for cycle 24 sunspots which followed the termination of the cycle 23 bands. Sunspots rapidly appear and grow on that band for several years until the next (oppositely-signed) band appears at high latitude. A relationship that is determined empirically in Paper I and defines the maximum activity level of that new cycle by triggering a downturn in sunspot production on the low-latitude band. The perpetual interaction of these temporally offset 22-year activity bands drives the quasi-11-year cycle of sunspots which forms the envelope of the Sun's magnetically driven activity where the degree of overlap in the bands appears to inversely govern our star's sunspot production (more overlap, less spots and vice versa). This observational evidence presented in Paper~I points to the rotational energy at the bottom of our Star's convective interior as being a/the major driver of its long-term (magnetic) evolution, where we suppose that perturbations to that system, and the sunspots that those perturbations produce have the potential to impact short-term variability.

\begin{figure}
\epsscale{1.00}
\plotone{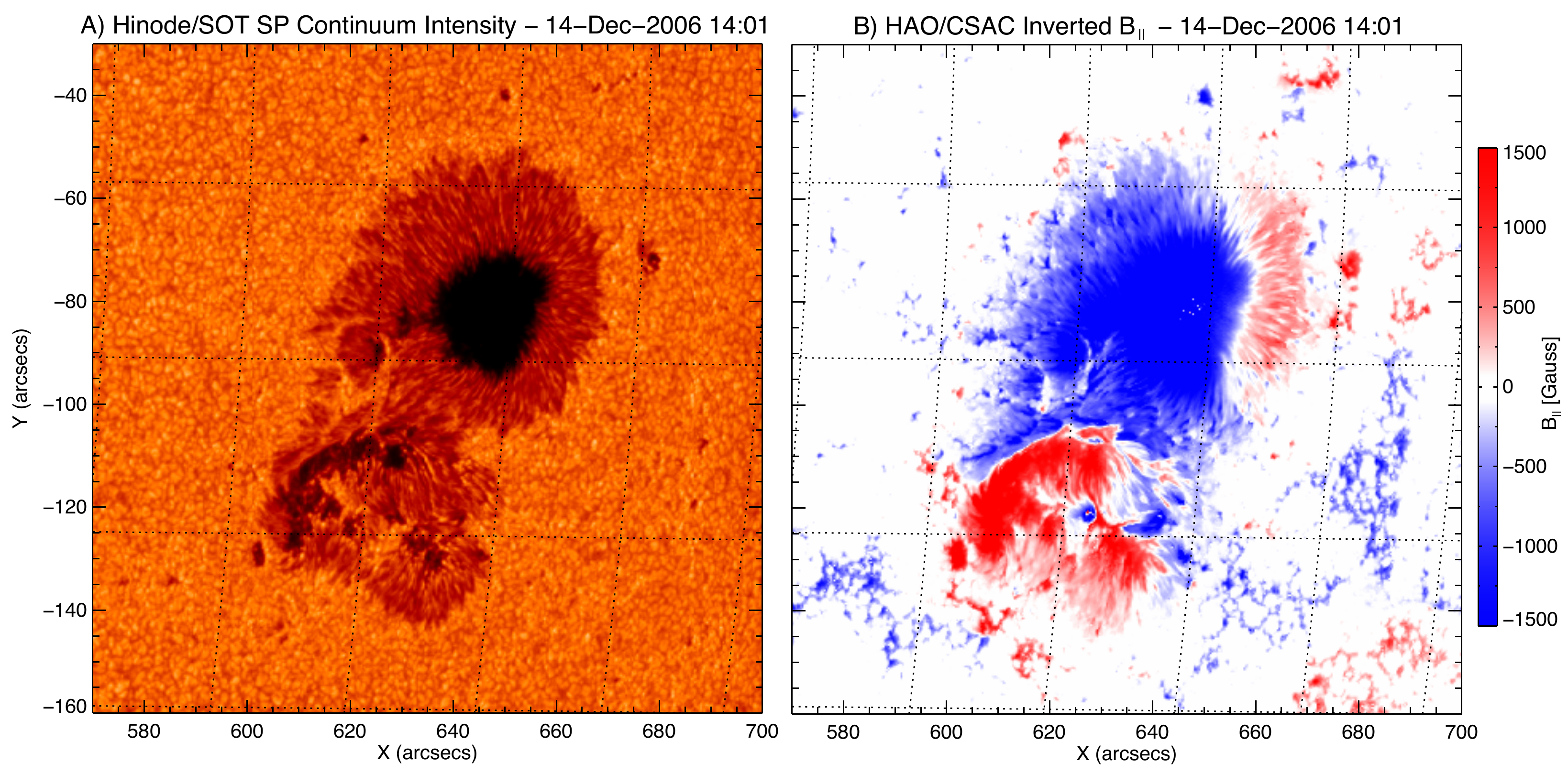}
\caption{\hinode/SOT Spectropolarimeter observations of NOAA AR 10930 a complex active region ($\beta\gamma\delta$---Mount Wilson; {\it Dki}---McIntosh Classification) near the solar equator on Dec 14, 2006. Panel A shows the continuum intensity neighboring the \ion{Fe}{1} 6302.5\AA{} line and B, the line-of-sight magnetogram inferred from a Milne-Eddington inversion of the measured Stokes Profiles. This active region fired the final significant salvos of solar cycle 23, including four X-class flares, as it traversed the Earth-facing side of the Sun. \label{f2}}
\end{figure}

In this Letter we present an extension of the activity band overlap and interaction concept that was introduced in Paper~I. We consider the formation of complex active regions (Fig~\pref{f2}) in light of this picture, where and when these regions might form and their potential to form delta regions---the subset of active regions responsible for the vast majority of M and X-Class flares \citep[e.g.,][]{2008LRSP....5....1B}. It is then a further extension to consider epochs when the most catastrophic flares that give rise to the most intense geomagnetic storms, i.e., those of a magnitude similar to the Sept 1 1859 ``Carrington Event'' \citep{1859MNRAS..20...13C}, have a higher probability of occurrence. 

\section{Band-Band Interaction}\label{bands}
\begin{figure}
\epsscale{1.00}
\plotone{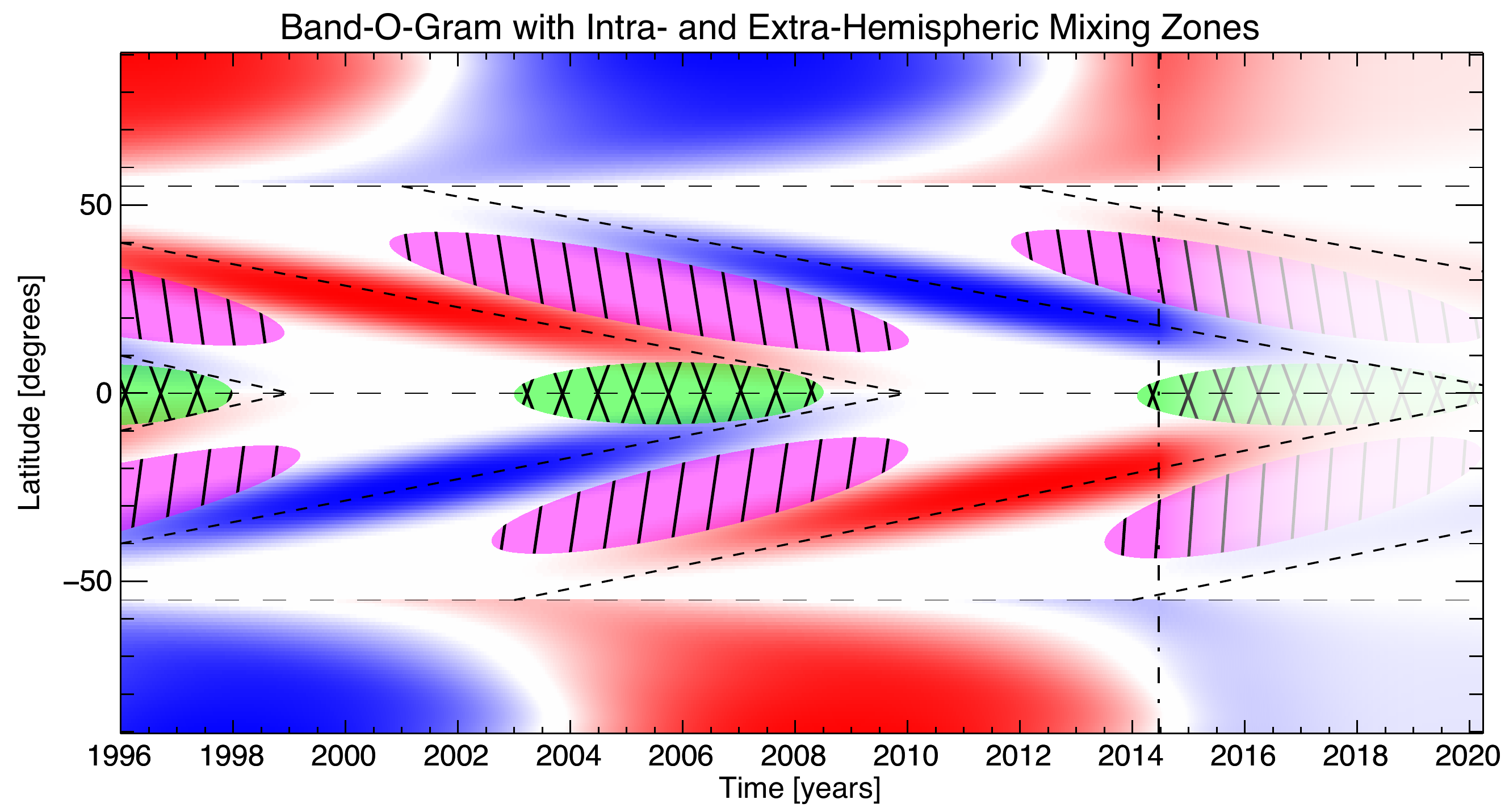}
\caption{The data-determined schematic of the migrating activity bands that belong to the 22-year magnetic activity cycle from \citet{McIntosh2014b}---compare with Fig.~\pref{f1}. In addition to the forecast we have now estimated the latitudes and times for which cross-band mixing becomes important---these regions are drawn as pink and green ellipses (see text). The main distinguishing factor between these two interaction regions is the impact that helicity can have on them and so the green ellipses around the equator are cross-hatched to indicate the potential for cross-band and opposite-helicity mixing across the solar equator. \label{f3}}
\end{figure}

Figure~\pref{f3} builds upon the activity bands of Fig.~\pref{f1} to illustrate the spatio-termporal zones where we would expect to see signatures of band-band interaction occur naturally, here drawn as pink and green hatched ovals. The pink ovals represent ``simple'' mixing. In those zones the erupting magnetic flux could come from either polarity of band and mix (in the interior) before emerging to create what would appear to be an ``anomalous'' active region---with characteristics of both the current {\em and\/} upcoming cycles, but this would be many years before the true spots and active region complexes of the latter appear. Also, the active region pairs at high latitude would have the same helicity given that the bands from which the flux originates are in the same hemisphere---left-handed in the north, right-handed in the south \citep{1941ApJ....93...24R,2010MNRAS.402L..30Z}. You will notice that this process can {\em only\/} occur after the new cycle bands at high-latitude have started their equatorward migration---by our earlier declaration this is {\em after\/} solar maximum in the declining phase of the sunspot cycle for each hemisphere. So, we would expect the frequency of mixed, or reversed, active region complexes to increase significantly in the declining phase---occurring on the high-latitude periphery of the butterfly wings (the tear-shaped areas enclosing the hemispheric sunspots of each cycle). From a space weather climatology standpoint, we would be able to watch these zones carefully for such flux systems and be able to monitor how they interact with their surroundings. 

The green cross-hatched ovals indicate a second zone, an equatorial mixing zone, where the bands of the dying cycle interact at low-latitudes across the equator. These bands will have opposite polarity {\em and\/} opposite helicity since they are anchored in different hemispheres and experience the different signs of the Coriolis force. These active region complexes would occur on the equatorward periphery of the butterfly wings in the deep descending phase of one sunspot cycle and into the early ascending phase of the next. We anticipate that the cross-polarity, opposite-helicity mixing of the low-latitude activity bands should produce violent active region complexes---as demonstrated by studies of flux system interaction \citep[e.g.,][]{2001ApJ...553..905L,2007A&A...466..367A}. Again, from a forecast standpoint we would know roughly where and when to look for such regions as they emerge.

\begin{figure}
\epsscale{1.00}
\plotone{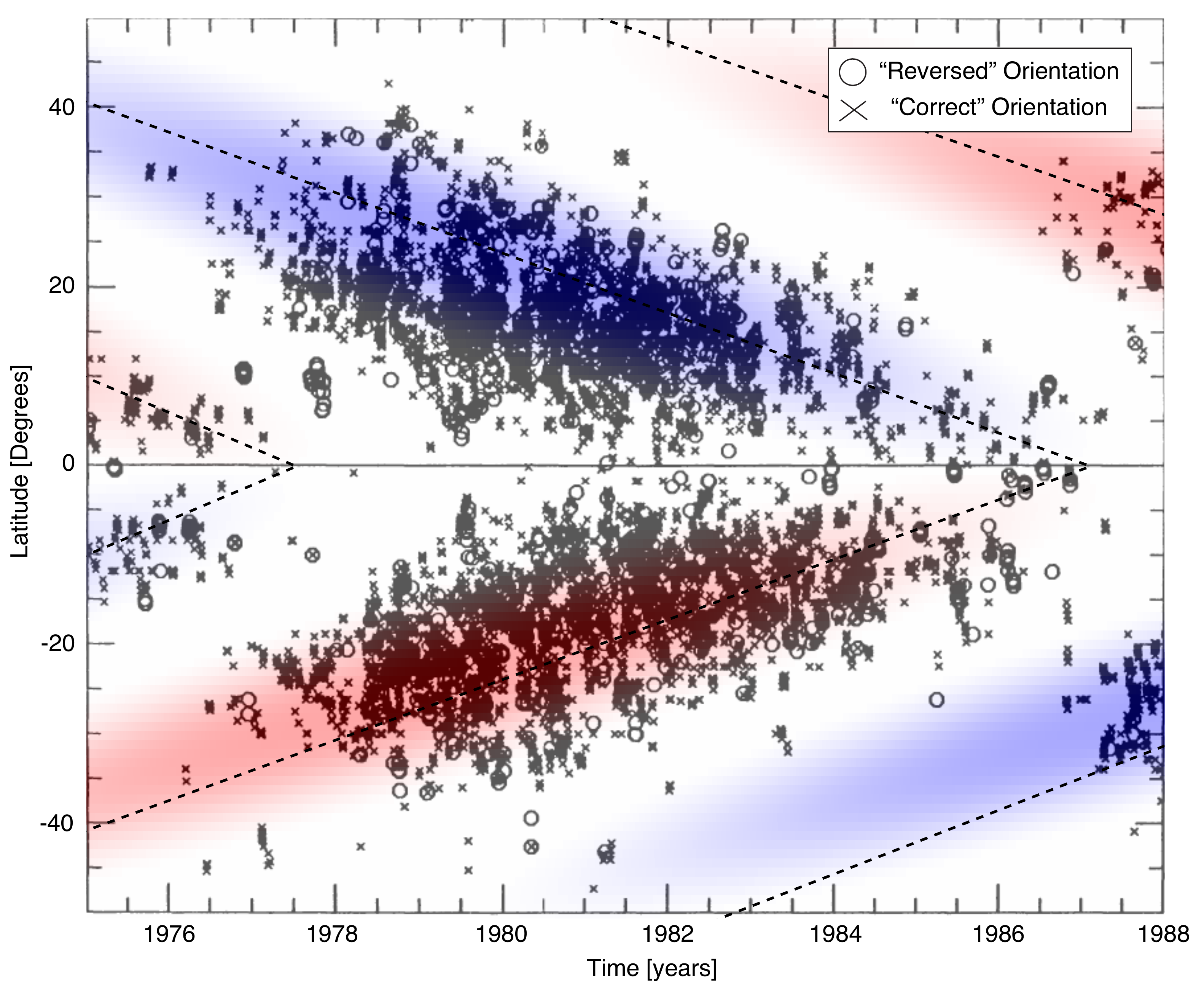}
\caption{Contrasting the orientation of magnetic bipolar regions in a latitude-time diagram of ``properly oriented'' (x) and ``reverse oriented'' ($\circ$) bipolar regions from sunspot cycles 20, 21, and 22 \citep{1992ASPC...27..335H}. Overlaid are estimates of the 22-year magnetic activity band locations that modulate those sunspot cycles (cf. Fig.~\pref{f1}). The extracted figure and data points of \citet{1992ASPC...27..335H} is reproduced with permission of PASJ. \label{f4}}
\end{figure}

Figure~\pref{f4} uses the results published by \cite{1992ASPC...27..335H} that were extracted from the detailed studies of the magnetic configurations of active regions in cycles 20, 21, and 22 by \citet{1989SoPh..123..271H, 1991SoPh..135...43H}. To guide the reader we have estimated the positions of the activity bands in this butterfly diagram using the same technique used to create Figs.~11 and 12 in section~5 of Paper~I. In short, we use the hemispheric sunspot maxima to determine the start time of band migration at high latitude and the abrupt increase of spot area (on the high-latitude new cycle bands) to estimate the time of termination for the low-latitude bands. Together, these landmarks allow us to linearly estimate where the activity band would be. You will notice that the vast majority of points in the plot (the crosses) have the ``correct orientation'' in that they appear ``as expected'' for that cycle to use the language of \cite{1992ASPC...27..335H}. The open circles, on the other hand, represent anomalous active regions, those where the footprints of the system have the ``reversed orientation'' from what was expected. Many of those open circles appear in locations that we would have expected given the interacting flux-band picture introduced in Paper~I and drawn in Fig.~\pref{f3} above.

\subsection{``Delta'' Regions}\label{deltas}
\begin{figure*}
\epsscale{1.00}
\plotone{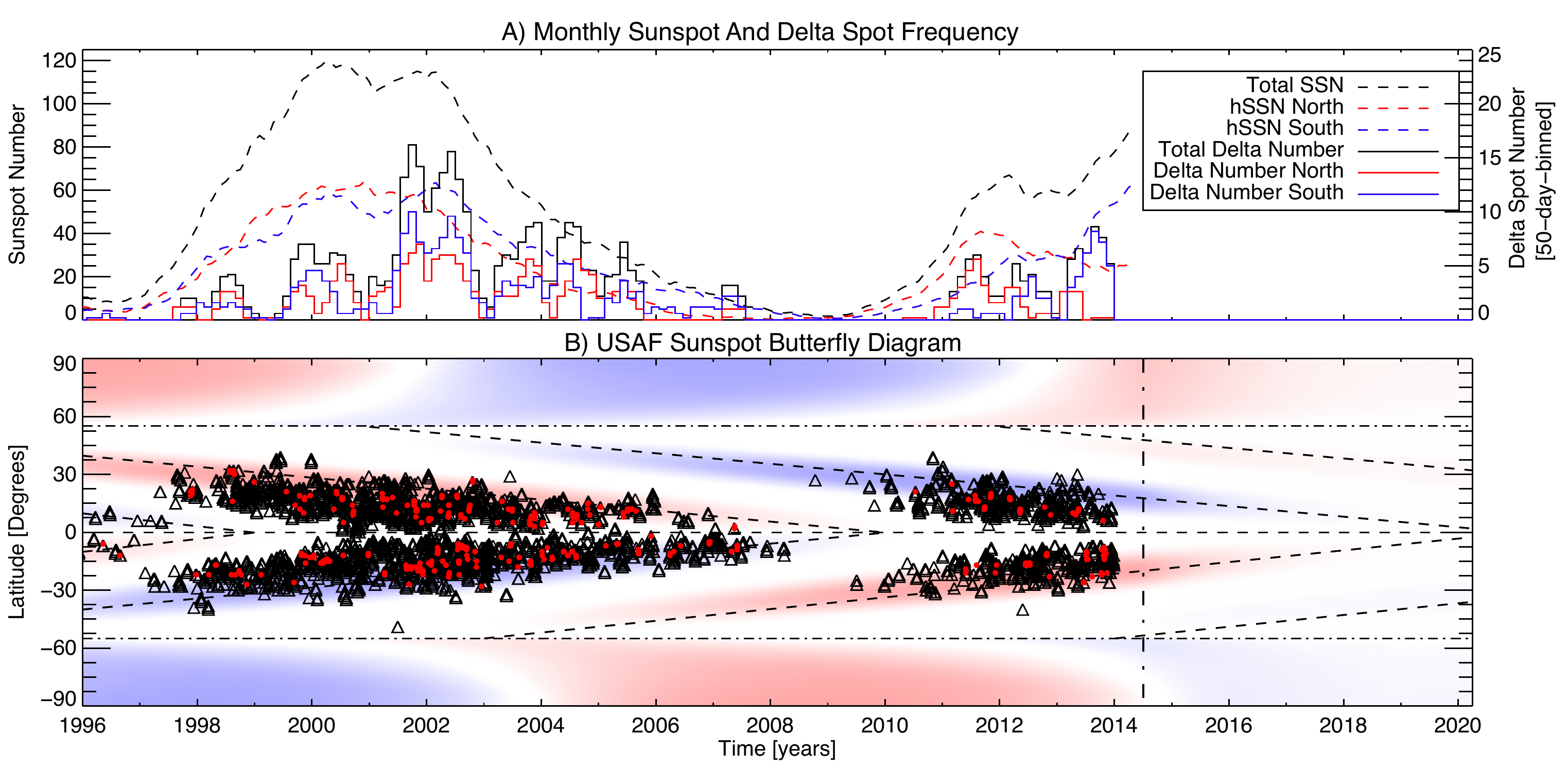}
\caption{A: Comparing the number of delta-classified regions (black solid line) with the hemispherically decomposed number of the same (red--north; blue--south) versus the corresponding total and hemispheric (monthly) sunspot numbers. B: comparing regular (black $\triangle$) and delta-classified regions (red $\circ$) locations for solar cycles 23 and 24 from the US Air Force active region record. These locations are overlaid on the activity bands shown in Figs.~\pref{f1} and~\pref{f3}. \label{f5}}
\end{figure*}

Returning to an analysis of contemporary data, Fig.~\pref{f5} presents a comparison between the activity band interaction schematic of Fig.~1, the hemispheric and total sunspot numbers (with corresponding numbers of delta regions), and the latitude-time distribution of active regions (black $\triangle$) and regions with a delta-classification (red $\circ$) locations taken from the US Air Force\footnote{The USAF sunspot record can be retrieved from Dr. David Hathaway at NASA/MSFC \url[http://solarscience.msfc.nasa.gov/greenwch.shtml]{http://solarscience.msfc.nasa.gov/greenwch.shtml}.} active region record. The textbook definition of a ``delta'' region as two oppositely signed umbrae (separated by less than 2 degrees) within one penumbra traces back to \citet{1919ApJ....49..153H} and \citet{1960AN....285..271K} who added the delta classification to the Hale's original $\alpha$, $\beta$ and $\gamma$ sunspot classifications. As discussed above, these delta regions, which compose only a few percent of the total number of sunspots, are the subset responsible for the vast majority of M and X-Class flares \citep[e.g.,][]{2008LRSP....5....1B}. Therefore, their origins are critically important in the context of space weather.

Comparing  Fig.~\pref{f5} with our earlier analysis and discussion of Fig.~\pref{f3} we see that the delta-classified regions appear throughout the active region butterfly diagram, but with a possible preference of a drift from the high latitude to low latitude periphery of the butterfly diagram. In the deepest part of the descending phase --- from $\sim$2004 on --- the delta-classified regions tend to appear predominantly on the low-latitude periphery of the butterfly in each hemisphere. These delta regions could be born of extra-hemispheric band interaction where bands of opposite polarity {\em and} opposite helicity can mix. Further, considering the total and hemispheric delta numbers for cycle 23 (panel A), we see that the number of delta-classified regions appears to be weighted to the descending phase of the cycle. We also notice an apparent clustering (in time) of regions which, when considered as a timeseries, create a quasi-periodicity in the delta region numbers of each hemisphere (and the total). The periodicity and timing of the peaks in the delta region timeseries match other proxies of magnetic solar activity that have been tied to significant surges of magnetic flux emergence from the deep interior that are of unknown physical origin \citep{McIntosh2014c}. An extended study of this pattern is beyond the scope of this Letter, but an effort in preparation \citep{LeamonInPrep} will use the Mt.\ Wilson activity and sunspot record \citep[e.g.,][]{Pevtsov2013} back to the early twentieth century to explore these relationships.

\section{Discussion}\label{discuss}
In the above analysis we have noted that the band-band interaction hypothesis of Paper~I can be used to explain the occurrence of some, but not all, reversed or irregular sunspot groups and active regions. The band-band interaction scheme that we've discussed above would lead us to deduce that the occurrence rate of such regions (in each hemisphere) should grow after the sunspot maximum (in each hemisphere). This is a time when there are two activity bands in each hemisphere. The interaction regions between the activity bands should be preferred locations for irregular active regions to form. This appears to be supported by the evolution shown in Fig.~\pref{f4} \citep[and][]{1992ASPC...27..335H}. The equatorial interaction region is of particular interest in the context of space weather. In this region it is possible that activity bands of mixed sign {\em and\/} helicity can interact with each other across the equator. The analysis of Fig.~\pref{f5} would tend to support that deduction. 

In addition to the declaration of irregular, reversed, or Hale's law-violating active regions \citep[][]{Hale1924}, there are numerous ``anomaly'' reports in the literature which would tend to support our hypothesis and lend greater support to the band-band interaction concept as a key framer of solar activity. These anomalies include very high latitude sunspot groups after solar maximum \citep[][just as the next bands appear]{1860MNRAS..20..254C}, regions of incorrectly sensed magnetic helicity showing up in the wrong hemisphere around the equator \citep{2010MNRAS.402L..30Z, 2010ApJ...719.1955Z, 2012MNRAS.419..799Z, 2014ApJ...783L...1L}, sunspots with incorrectly oriented super-penumbral whorls \citep[e.g.,][]{1941ApJ....93...24R,1996MNRAS.278..821P,2004ApJ...608.1148B} and X-ray sigmoids of the wrong shape in the wrong hemisphere \citep[e.g.,][]{Rust1996, 2002JGRA..107.1234L,2007ApJ...671L..81C}. 

Indeed, while neither \cite{Rust1996} or \cite{2002JGRA..107.1234L} found lockstep agreement between hemisphere and sigmoid helicity handedness, the greater preponderance of ``correctly'' kinked sigmoids ($\sim$4 in 5 by the former, compared to ``only'' $\sim$2 in 3 to the latter) can be explained by when the two studies' data were collected.\citet{Rust1996} focused on just 15~months (October 1991--January 1993) at solar maximum when our band hypothesis would imply there is only the single flux system present, whereas  \citeauthor{2002JGRA..107.1234L}'s dataset covered the entire descending phase of cycle~22, and early rise phase of cycle~23, when, as we have seen, band-band interactions would favor the appearance of mixed-helicity active regions.
However, a more detailed study is required to identify such regions in the full historical record and investigate if these reported anomalies really are a very natural by-products of this complex band-band interaction taking place in the solar interior.

\subsection{The Origin of Extreme (Space Weather) Events?}\label{extremes}
Extending our deductive reasoning we consider the occurrence of the most profound of solar events. These solar storms \citep[e.g.,][]{1859MNRAS..20...13C} would appear at first glance to occur randomly---the vast majority do not occur at sunspot maxima \citep[e.g.,][]{2006AdSpR..38..280O}.
Based on our earlier deduction that (devastatingly) complex active regions can occur in the equatorial interaction region where cross-polarity and opposite-helicity bands can interact we would naturally expect a bias to times of activity band overlap---the deep declining phase and early ascending phase of solar activity (cf. Fig.~1). Therefore, we deduce that the length of time spent by the activity bands at very low-latitudes is strongly related to the probability that a devastating solar storm will occur. Paper~I discussed that the number of sunspots produced in a given cycle was inversely related to the degree of overlap between the interacting bands. Shorter overlap times tend to produce a long ascending phase, but a short and strong cycle, while longer overlap times produce shorter ascending phases, and weaker, longer cycles. \citet{McIntosh2014d} discusses a mode where the overlap of the activity bands is at its greatest possible extent---one where there may be no sunspots produced for an extended period of time---a grand (sunspot) minimum---even though the Sun would continue to cycle. It is very likely that these {\em very\/} lengthy overlaps, times of prolonged lows in sunspot numbers are {\em significantly more\/} likely to produce an event of the scale of that witnessed by \citet{1859MNRAS..20...13C}. Unfortunately, the full record of such extreme events is limited by the fact we ``catch'' only those that are geo-effective, thus creating a selective bias in the sample required to fully test of our hypothesis. The forthcoming study of \citet{{LeamonInPrep}} will consider this topic at greater length as it has considerable potential impact on our ability to foretell of potentially cataclysmic solar storms.

\section{Conclusion}\label{conclusion}
The discussion presented above logically extends the concept presented in Paper~I of how the activity bands that belong to the 22-year magnetic activity cycle (phenomenologically) explain the landmarks of the sunspot cycle to complex active regions and delta regions. We deduce that such complex regions form as a consequence of the intra- and extra-hemispheric interactions taking place in the solar interior between those bands---this interaction can naturally result in active regions of mixed polarity and built-in complexity before emergence into the outer atmosphere. However, we acknowledge that many physical processes internal to the activity band could also produce ``irregular'' sunspots and active regions \citep[e.g.,][]{2007A&A...466..367A, Pevtsov2013, 2014ApJ...783L...1L}. Further, we have deduced that a class of such complex active regions can exhibit interaction from flux systems of mixed helicity in addition to polarity. Those regions most frequently would occur in the deep descending phase of sunspot cycles and present significant threat to much of the solar system as they can explosively release their (stored) magnetic energy \citep{2001ApJ...553..905L} in close proximity to the solar equator. When these near-equatorial bands remain in close proximity for extended times, as one may expect on the entry into grand minima \citep{McIntosh2014d}, the likelihood of a profound eruptive event would increase.

We acknowledge that while compelling, this is but one way to generate complex active regions and delta spots. Another way is that new flux tubes from the same band, but offset in latitude or longitude (or depth), could be sufficiently buffeted by turbulent motions deep below the photosphere so as to emerge into existing systems giving the appearance of an incorrectly oriented system.

Nevertheless, the hypothesis presented here points to a picture of the Sun's deep convective interior, and the magnetic interactions taking place therein, that is not yet accessible to numerical simulation although we note that steps in that direction have been taken \citep[e.g.,][]{Weber2012, 2014SoPh..289..441N}. Such investigations are necessary to develop the considerable forecast skill required to protect Earth- and Space-based infrastructure required for our technologically-driven society.

\acknowledgements
NCAR is sponsored by the National Science Foundation. We acknowledge support from NASA contracts NNX08BA99G, NNX11AN98G, NNM12AB40P, NNG09FA40C ({\em IRIS}), and NNM07AA01C ({\em Hinode}). Hinode SOT/SP Inversions were conducted at NCAR under the framework of the Community Spectro-polarimtetric Analysis Center (\url[http://www.csac.hao.ucar.edu/]{CSAC}). We also thank Alan Title, Giuliana de Toma. Fraser Watson, and Luca Bertello for the discussion that form the basis of a deeper (historical) investigation to follow.


\begin{thebibliography}{}

\bibitem[Archontis et al.(2007)]{2007A&A...466..367A} 
Archontis, V., Hood, A.~W., \& Brady, C.\ 2007, \aap, 466, 367 

\bibitem[Balasubramaniam et al.(2004)]{2004ApJ...608.1148B} 
Balasubramaniam, K.~S., Pevtsov, A., \& Rogers, J.\ 2004, \apj, 608, 1148 

\bibitem[Beck et al.(1998)]{1998Natur.394..653B} 
Beck, J.~G., Duvall, T.~L., \& Scherrer, P.~H.\ 1998, \nat, 394, 653 

\bibitem[Benz(2008)]{2008LRSP....5....1B} 
Benz, A.~O.\ 2008, Living Reviews in Solar Physics, 5, 1 

\bibitem[Canfield et al.(2007)]{2007ApJ...671L..81C} 
Canfield, R.~C., Kazachenko, M.~D., Acton, L.~W., et al.\ 2007, \apjl, 671, L81 

\bibitem[Carrington(1859)]{1859MNRAS..20...13C} 
Carrington, R.~C.\ 1859, \mnras, 20, 13 

\bibitem[Carrington(1860)]{1860MNRAS..20..254C} 
Carrington, R.~C.\ 1860, \mnras, 20, 254 

\bibitem[Golub et al.(1974)]{1974ApJ...189L..93G} 
Golub, L., Krieger, A.~S., Silk, J.~K., Timothy, A.~F., \& Vaiana, G.~S.\ 1974, \apjl, 189, L93 

\bibitem[Hale et al.(1919)]{1919ApJ....49..153H} 
Hale, G.~E., Ellerman, F., Nicholson, S.~B., \& Joy, A.~H.\ 1919, \apj, 49, 153

\bibitem[{{Hale}(1924)}]{Hale1924} 
Hale, G.E., Proc Natl Acad Sci U S A. Jan 1924; 10(1): 53?55.

\bibitem[Harvey(1992)]{1992ASPC...27..335H} 
Harvey, K.~L.\ 1992, Ast. Soc. Pac. Conf. Ser., 27, 335 

\bibitem[Hathaway et al.(2013)]{2013Sci...342.1217H} 
Hathaway, D.~H., Upton, L., \& Colegrove, O.\ 2013, Science, 342, 1217

\bibitem[Howard(1989)]{1989SoPh..123..271H} 
Howard, R.~F.\ 1989, \solphys, 123, 271 


\bibitem[Howard(1991)]{1991SoPh..135...43H} 
Howard, R.~F.\ 1991, \solphys, 135, 43 

\bibitem[K{\"u}nzel(1960)]{1960AN....285..271K} 
K{\"u}nzel, H.\ 1960, Astronomische Nachrichten, 285, 271

\bibitem[Leamon et al.(2002)]{2002JGRA..107.1234L} 
Leamon, R.~J., Canfield, R.~C., \& Pevtsov, A.~A.\ 2002, Journal of Geophysical Research (Space Physics), 107, 1234

\bibitem[{{Leamon} {et~al.}(2015)}]{LeamonInPrep}
Leamon, R.~J., McIntosh, S.~W., Watson, F.~T., de Toma, G., McKenzie, D.~E., Tiitle, A.~M., 2015 \apj, in prep.

\bibitem[Linton et al.(2001)]{2001ApJ...553..905L} 
Linton, M.~G., Dahlburg, R.~B., \& Antiochos, S.~K.\ 2001, \apj, 553, 905 

\bibitem[Liu et al.(2014)]{2014ApJ...783L...1L} 
Liu, Y., Hoeksema, J.~T., \& Sun, X.\ 2014, \apjl, 783, L1 

\bibitem[McIntosh(1992)]{1992ASPC...27...14M} 
McIntosh, P.~S.\ 1992, The Solar Cycle, 27, 14 

\bibitem[McIntosh et al.(2007)]{2007ApJ...654..650M} 
McIntosh, S.~W., Davey, A.~R., Hassler, D.~M., et al.\ 2007, \apj, 654, 650

\bibitem[{{McIntosh} {et~al.}(2014a)}]{McIntosh2014a} 
McIntosh, S.~W., Wang, X., Leamon, R.~J., \& Scherrer, P.~H.\ 2014a, \apjl, 784, L32 

\bibitem[{{McIntosh} {et~al.}(2014b)}]{McIntosh2014b}
McIntosh, S.~W., et~al., 2014b \apj, 792, 12

\bibitem[{{McIntosh} {et~al.}(2014c)}]{McIntosh2014c}
McIntosh, S.~W., Leamon, R.~J, Krista, L. D., Title, A.~M. Hudson, H.~S., Riley, P., Harder, J.~W. , Kopp, G., Snow, M., Woods, T.~N., Kasper, J.~C., Stevens, M.~L., Ulrich, R.~K., 2014c \url[https://dl.dropboxusercontent.com/u/3057160/McIntoshSWx_Submit.pdf]{"On the Quasi-Periodic Forcing of the Sun's Eruptive, Radiative, and Particulate Output"}, Nature Communications, in press

\bibitem[{{McIntosh} {et~al.}(2014d)}]{McIntosh2014d}
McIntosh, S.~W., Leamon, R.~J, Centeno-Elliott, R., 2014c, ``Deciphering Solar Magnetic Activity: Grand Activity Maxima'', Frontiers in Space Science, submitted (preprint available on request).

\bibitem[Nelson et al.(2014)]{2014SoPh..289..441N} 
Nelson, N.~J., Brown, B.~P., Sacha Brun, A., Miesch, M.~S., \& Toomre, J.\ 2014, \solphys, 289, 441 

\bibitem[Odenwald et al.(2006)]{2006AdSpR..38..280O} 
Odenwald, S., Green, J., \& Taylor, W.\ 2006, Advances in Space Research, 38, 280 

\bibitem[{{Pevtsov} {et~al.}(2013)}]{Pevtsov2013}
Pevtsov, A.~A., Bertello, L., Tlatov, A.~G., Kilcik, A., Nagovitsyn, Y~.A., Cliver, E. W.,  2013, \solphys, XXX, YYY

\bibitem[Peter(1996)]{1996MNRAS.278..821P} 
Peter, H.\ 1996, \mnras, 278, 821 

\bibitem[Richardson(1941)]{1941ApJ....93...24R} 
Richardson, R.~S.\ 1941, \apj, 93, 24 

\bibitem[{{Rust} \& {Kumar}(1996)}]{Rust1996}
Rust, D.~M., Kumar, A, 1996, \apj, 464, L199

\bibitem[{{Weber} {et~al.}(2012){Weber}, {Fan}, \& {Miesch}}]{Weber2012}
{Weber}, M.~A. and {Fan}, Y. and {Miesch}, M.~S., 2012, \solphys, 287, 239

\bibitem[Wilson et al.(1988)]{1988Natur.333..748W} 
Wilson, P.~R., Altrock, R.~C., Harvey, K.~L., Martin, S.~F., \& Snodgrass, H.~B.\ 1988, \nat, 333, 748 

\bibitem[Zhang et al.(2010a)]{2010MNRAS.402L..30Z} 
Zhang, H., Sakurai, T., Pevtsov, A., et al.\ 2010, \mnras, 402, L30 

\bibitem[Zhang et al.(2010b)]{2010ApJ...719.1955Z} 
Zhang, H., Yang, S., Gao, Y., et al.\ 2010, \apj, 719, 1955 

\bibitem[Zhang(2012)]{2012MNRAS.419..799Z} 
Zhang, H.\ 2012, \mnras, 419, 799 
\end{thebibliography}
\end{document}